\documentclass[twocolumn]{aastex63}

\usepackage[utf8]{inputenc}
\usepackage{hyperref}

\newcommand{\angstrom}{~\textup{\AA}}

\received{\today}
\revised{\today}
\accepted{some day}
\submitjournal{ApJ}

\shorttitle{Asymmetry revisited}
\shortauthors{Yuan et al.}

\begin{document}

\title{Asymmetry Revisited: The Effect of Dust Attenuation and Galaxy Inclination}

\correspondingauthor{Fang-Ting Yuan}
\email{yuanft@shao.ac.cn}

\author{Fang-Ting Yuan}
\affiliation{Key Laboratory for Research in Galaxies and Cosmology, Shanghai Astronomical Observatory, Chinese Academy of Sciences, 80 Nandan Road, Shanghai 200030, China}

\author{Jiafeng Lu}
\affiliation{Key Laboratory for Research in Galaxies and Cosmology, Shanghai Astronomical Observatory, Chinese Academy of Sciences, 80 Nandan Road, Shanghai 200030, China}
\affiliation{University of Chinese Academy of Sciences, No. 19A Yuquan
Road, Beijing 100049, China}

\author{Shiyin Shen}
\affiliation{Key Laboratory for Research in Galaxies and Cosmology, Shanghai Astronomical Observatory, Chinese Academy of Sciences, 80 Nandan Road, Shanghai 200030, China}

\author{M\'{e}d\'{e}ric Boquien}
\affiliation{Universidad de Antofagasta, Centro de Astronom\'{i}, a
Avenida Angamos 601, Antofagasta 1270300, Chile}

\begin{abstract}
Dust attenuation of an inclined galaxy can cause additional asymmetries in observations, even if the galaxy has a perfectly symmetric structure. {Taking advantage of the integral field spectroscopic data observed by the SDSS-IV MaNGA survey, we investigate the asymmetries of the emission-line and continuum maps of star-forming disk galaxies.} We define new parameters, $A_a$ and $A_b$, to estimate the asymmetries of a galaxy about its major and minor axes, respectively. Comparing $A_a$ and $A_b$ in different inclination bins, we attempt to detect the asymmetries caused by dust. For the continuum images, we find that $A_a$ increases with the inclination, while the $A_b$ is a constant as inclination changes. Similar trends are found for $g-r$, $g-i$ and $r-i$ color images. The dependence of the asymmetry on inclination suggests a thin dust layer with a scale height smaller than the stellar populations. For the H$\alpha$ and H$\beta$ images, neither $A_a$ nor $A_b$ shows a significant correlation with inclination. Also, we do not find any significant dependence of the asymmetry of $E(B-V)_g$ on inclination, implying that the dust in the thick disk component is not significant. Compared to the SKIRT simulation, the results suggest that the thin dust disk has an optical depth $\tau_V\sim0.2$. This is the first time that the asymmetries caused by the dust attenuation and the inclination are probed statistically with a large sample. Our results indicate that the combination of the dust attenuation and the inclination effects is a potential indicator of the 3D disk orientation.

\end{abstract}


\section{Introduction}
The morphology and the structures of galaxies provide us with important clues for understanding the physics of galaxy formation. Historically, the morphology and the structures of galaxies were determined {by visual inspection \citep[e.g.,][]{hubble1926}}. With the advent of photometry and the use of CCD, a large amount of work has been done to measure the morphology and structure of galaxies quantitatively \citep[e.g.,][and reference therein]{devaucouleurs1948, sandage1961, kormendy1977, peng2002,simard2011,conselice2014}. 


The techniques to measure the galaxy morphology can be separated into the parametric method and the non-parametric method. Compared with the parametric method, the non-parametric method does not depend on the assumption of analytical forms of galaxy light distribution. The most common non-parametric method used at present is through the CAS system, which measures the concentration (C), asymmetry (A), and clumpiness (S) of the galaxy light distribution \citep[e.g.,][]{abraham1994,schade1995,abraham1996,conselice2000,conselice2003,papovich2003,law2007,conselice2014}. Based on this system, \citet{lotz2004} added another two quantities, $M_{20}$ and Gini coefficient, to better describe the galaxy morphology. These parameters are ideal for deriving galaxy evolution over many epochs because they are designed to capture the major features of the underlying structures of galaxies in a way that does not depend on any assumed underlying form (as is done with the parametric method) and therefore can be measured out to high redshifts. 

In the CAS system, $A$ estimates the rotational asymmetry of galaxies. It is an indicator of what fraction of the light in a galaxy is in non-symmetric components and whether a galaxy is interacting with another galaxy \citep{conselice2006}. However, it is well known that dust can produce dramatic changes in morphology, which affects the measurements of asymmetry and other parameters \citep[e.g.][]{taylor-mager2007,lotz2008}. 

\begin{figure}
    \centering
    \includegraphics[width=\linewidth]{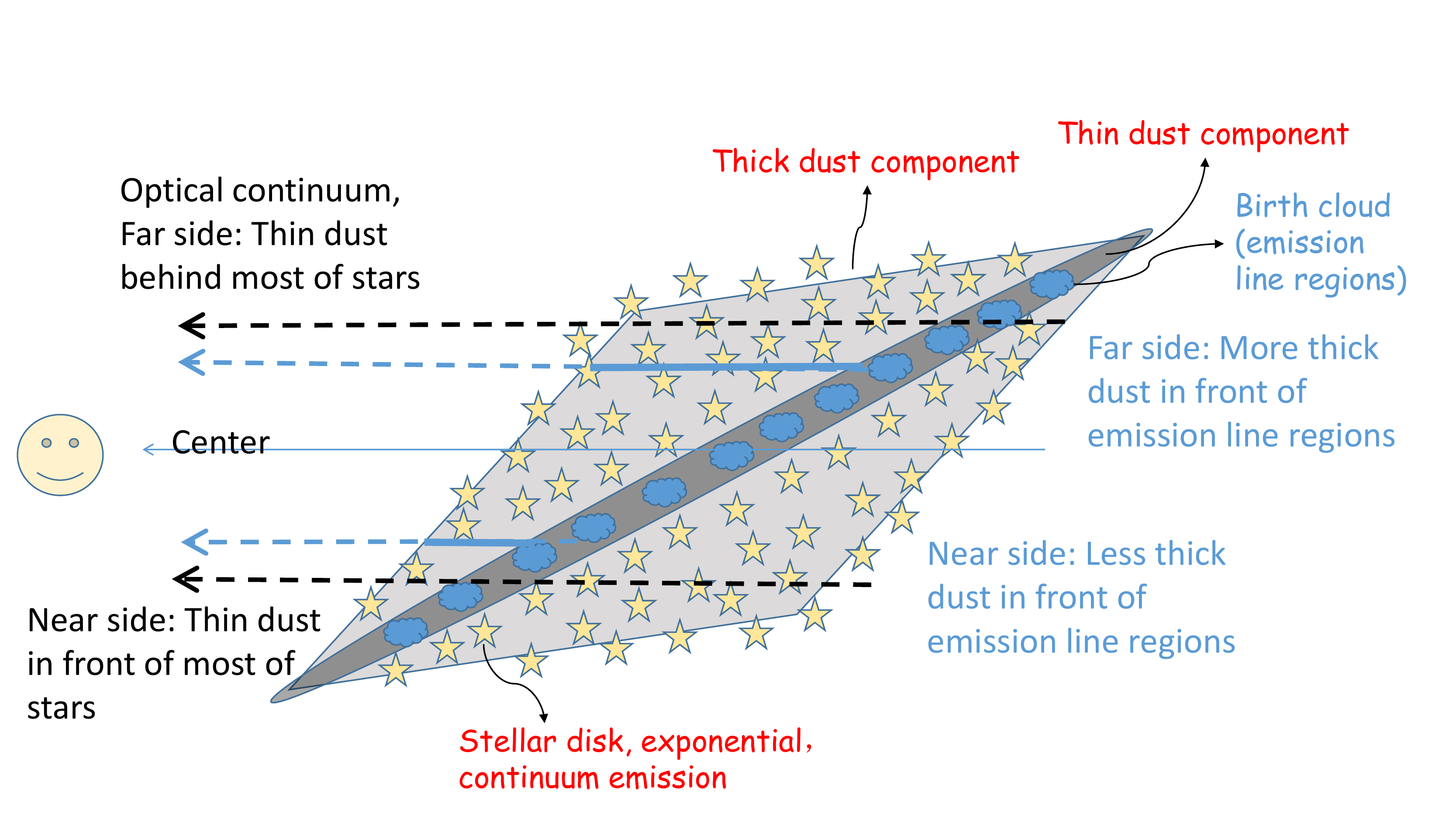}
    \caption{Illustration of a galaxy model with thin and thick dust components. The model contains a thick component with a small optical depth and a thin disk component with a relatively high optical depth. The birth cloud distributes as clumps in the thin disk. The recombination lines (H$\alpha$ and H$\beta$) originate from these clumpy birth clouds. The dustless bulge has been omitted for clarity.}
    \label{fig:illu}
\end{figure}

Even a simple dust distribution can affect the measurement of galaxy morphology. For example, a thin dust layer is presented in the mid-plane of a disk galaxy. Although the structure of the galaxy is perfectly symmetric, when the galaxy is inclined, we observe asymmetric light from the near and the far side of the galaxy. 
However, this asymmetry is only an observational effect, and therefore should be separated from the structural asymmetry. In reality, the dust distribution in galaxies is far more complicated than a thin disk. As shown in the commonly adopted two-component dust models (Figure \ref{fig:illu}), there is the dust in the birth clouds distributed in the center plane of galaxies, and the {ISM dust distributed more diffusely \citep[e.g.,][]{cf2000,tuffs2004,salim2007,conroy2009,chevallard2013}.} The dust components also have systematically different geometrical relations to stellar populations of different ages \citep[e.g.,][]{silva1998,cf2000,wg2000,popescu2000,popescu2011,yuan2018}. Such complications make it difficult to model the asymmetry caused by dust. 

The asymmetry caused by dust depends on the galaxy's inclination. {Even with no dust present}, the inclination can cause changes in the measurements of galaxy properties \citep{giovanelli1994,mollenhoff2006}. Recent works of \citet{devour2017,devour2019} used SDSS galaxies to discuss the inclination effects on the measured observational properties. The inclination effect is also used to test the opacity of galaxies, known as the Holmberg test \citep{holmberg1958,holmberg1975}. Many works have been dedicated to discuss the observational effect caused by inclination and dust \citep[e.g.,][]{tuffs2004,shao2007,graham2008,yip2010,wild2011}. 

Historically, the asymmetry caused by dust and inclination has been used to measure the galaxy tilt \citep[i.e., the direction of disk inclination, e.g.,][]{hubble1943,devaucouleurs1958,buta2003}, considering that the far side of a galaxy is less blocked by the dust lane when the galaxy inclined. The asymmetry has also been used by \cite{walterbos1988} to probe the dust attenuation in the Andromeda galaxy. However, previous works are limited to a few nearby galaxies. The validity of the method has never been tested for a large sample.


It is challenging to analyze the complicated observational effects caused by the dust and inclination in the studies of galaxy structures. The ability of the broadband photometric observation is limited because it is not sensitive to the dust attenuation in different components of galaxies. Integral field spectroscopy (IFS) may provide us with an unprecedented chance to examine these effects because it can obtain 3D data with the information of light distributions for different components of galaxies \citep{cappellari2016}. 

In this work, we present the first attempt to investigate the asymmetry caused by dust for different components of galaxies using a large sample of disk galaxies observed with IFS. Our data come from the Sloan Digital Sky Survey-IV (SDSS-IV) Mapping Nearby Galaxies at Apache Point Observatory (ManGA) survey \citep{bundy2015}. MaNGA provides IFS data for a large sample of galaxies for the first time, allowing for statistically meaningful studies of morphology in a wavelength range of 3600-10300\AA. Compared with SDSS single fiber survey, MaNGA gives us access to spatially resolved data extending to the outer regions of galaxies. MaNGA can map out information out to larger than 1.5 effective radius of its sample with a spatial resolution of 2.5 arcsecond \citep{yan2016}, sufficient for us to examine the asymmetries of galaxies.
The wavelength range covers all the strong nebular lines, including H$\alpha$ and H$\beta$, which are necessary to examine the asymmetries of the components that reside in the thin disks of galaxies. {The emission lines observed by MaNGA also allow us to obtain the spatially resolved BPT \citep*{bpt} excitation diagnostic diagram, which is crucial to disentangle and exclude non-star-forming regions before measuring the dust attenuation using Balmer decrements.}


This paper is organized in the following manner: We present our data and sample in Section \ref{sec:sample} and the method to investigate the dust asymmetry in Section \ref{sec:method}. Our main results are given in Section \ref{sec:results}. In Section \ref{sec:dustgeo}, we compare the results with the galaxy model to investigate the dust-star geometry and galaxy opacity. In Section \ref{sec:discussion}, we discuss the observational effects of the asymmetry and the uncertainties of this work. We summarize our work in Section \ref{sec:summary}. Throughout this paper, the redshifts and stellar masses are taken from the Nasa Sloan Atlas catalog (NSA; \citealt{blanton2005,wake2017}). The effective radius ($R_e$) {and the axis ratio $b/a$} of each galaxy is measured from Sloan Digital Sky Survey (SDSS) photometry by performing a S\'{e}rsic fit in the $r$ band. The inclination ($i$) is taken from the catalog of \citet{simard2011}. 

\section{Data and Sample}
\label{sec:sample}

SDSS-IV MaNGA is an IFS survey that has observed $\sim$10,000 galaxies with a median redshift of 0.03 using the BOSS spectrographs \citep{smee2013,drory2015} on the 2.5-meter SDSS telescope \citep{gunn2006,blanton2017}. MaNGA is equipped with fiber-bundle integral field units that vary in diameter from 12'' (19 fibers) to 32'' (127 fibers). These fibers are fed into the two dual-channel spectrographs, providing simultaneous wavelength coverage over 3600-10300 {\angstrom} at a spectral resolution $R\sim2000$. The angular resolution (the FWHM of the fiber-convolved PSF) of MaNGA data is about $2.5''$ \citep{yan2016}. In addition to a robust data-reduction pipeline \citep[DRP,][]{law2016}, MaNGA has developed a data-analysis pipeline (DAP) that provides higher-level data products \citep{westfall2019}.

\subsection{Sample}
\label{subsec:sample}
In this paper, we select our sample from MaNGA data release Mpl-8, which provide us with observations and data products for 6430 galaxies. In order to analyze the maps of emission lines, we first narrowed the sample to galaxies for which there are enough pixels in the emission line maps.  We require that 50\% spaxels of each selected galaxy have the H$\alpha$ flux with a detection larger than 5$\sigma$, which ensures that the emission line maps have enough valid spaxels for us to analyze the asymmetry. {This selection leaves us with a sample of 2784 sources.}  
 
We further require the S\'{e}rsic $n$ of the galaxies to be smaller than 2.5 to constrain our sample to disk dominated galaxies. Also, we limit the stellar mass range of our sample to be $10^{9.5}$ to $10^{11} M_{\odot}$. In such a mass range, galaxies are well evolved to develop stable disks. The final sample contains 1320 galaxies.

The sample contains galaxies that are dominated by the active galactic nuclei (AGNs) photoionisation and low-ionization (nuclear) emission-line regions [LI(N)ERs]. However, we find that the asymmetry of the maps is hardly affected by these phenomena. To examine the influence of AGNs and LI(N)ERs, {we further check our sample using BPT diagram \citep*{bpt}}. The MaNGA data allow us to examine the main ionization source spaxel-per-spaxel \citep{belfiore2016}. We find that there are 1095 galaxies with more than 50\% of the spaxels in 1$R_e$ that are classified as star-forming spaxels, which are those falling in the \citet{kewley2006} star-forming region in the [SII] BPTs, and the \citet{kauffmann2003} star-forming region in the [NII] BPT. We use these star-forming dominant galaxies for the analysis of emission-line maps. Including the AGN or LI(N)ER dominant galaxies in our sample does not affect our results.

\begin{figure}
    \centering
    \includegraphics[width=0.9\linewidth]{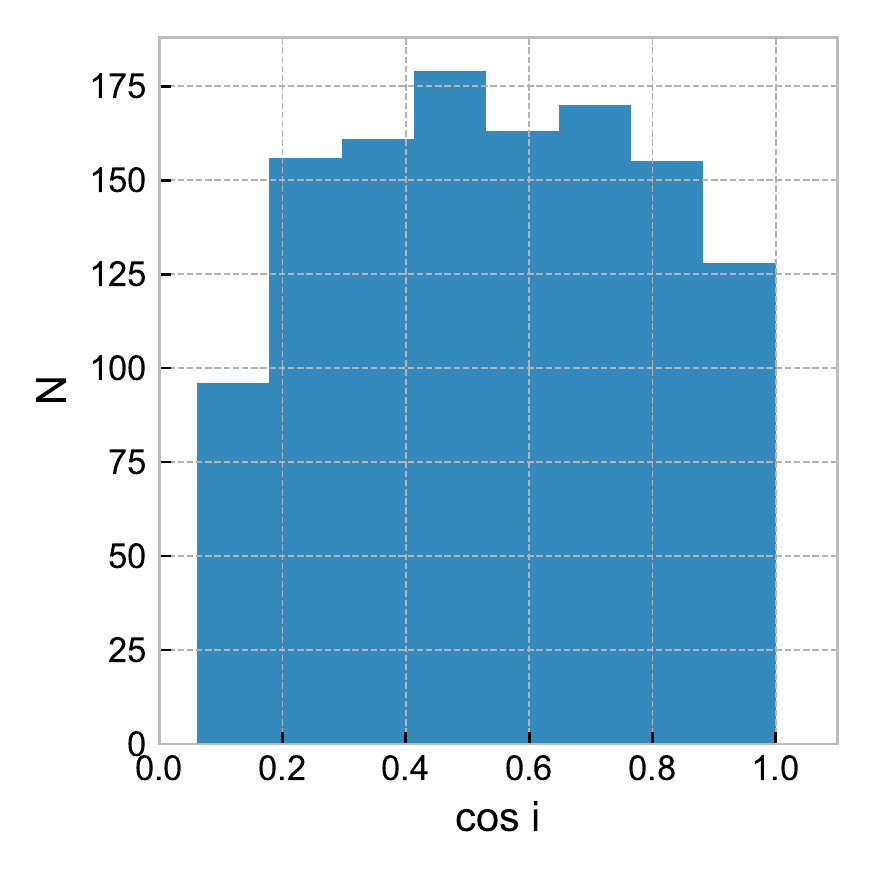}
    \caption{Distribution of inclination angle (i) of galaxies in our sample.}
    \label{fig:btoa}
\end{figure}

We examine the distribution of the inclination for our sample in Figure \ref{fig:btoa}. The distribution shows an average $\cos i\approx 0.5$, consistent with the expected average inclination of a random sample of galaxies with no selection effect at different inclination angles. 

We also examine the stellar masses of our galaxies at different inclinations, because the dust attenuation is also correlated with $M_*$ \citep{xu2007,martin2007,buat2009,heinis2014,am2016,bogdanoska2020}. We find that the average $\log M_{*}$ in each inclination bin varies from $10$ to $10.1$ dex, which is insignificant to affect dust attenuation. 

 Dust attenuation is also related to the metallicity \citep{brinchmann2004}. However, the metallicity plays an insignificant role in this work because we are focusing on the asymmetry of galaxy maps. {The only effect the metallicity can cause is the selection effect in each inclination bin, which is correlated with the stellar mass effect that we discussed above. The effect is also insignificant because the average metallicity varies less than 0.05 dex in each inclination bin according to the mass-metallicity relation \citep{tremonti2004}.}

\subsection{Emission line maps}

For this work, we use H$\alpha$ and H$\beta$ to calculate the asymmetry of emission line maps. We construct the emission line maps of galaxies using MaNGA {Mpl-8} data release, 
  including the output of the MaNGA data analysis pipeline (DAP, \citealt{westfall2019,belfiore2019}) for {6430 galaxies}. {This pipeline utilizes the code pPXF \citep{ce2004,cappellari2017} to fit the stellar continua and emission lines for galaxy spectra.} The flux measurements we adopt are obtained by Gaussian fitting for the emission lines. We use non-binned results from the DAP to ensure the best spatial resolution for the emission line maps. {These results are from the analysis of each individual spaxel. Each spaxel we used has a valid stellar continuum fit and emission-line fit.} 

 The spaxels with bad qualities are removed according to the mask arrays given by the DAP. We keep only the star-forming spaxels in H$\alpha$ and H$\beta$ maps according to the [NII] and [SII] BPT diagram with a minimum signal-to-noise ratio (S/N) of 3, as described in Section \ref{sec:sample}. Keeping spaxels that are not star-forming (i.e., AGNs, LI(N)ERs or composite) in these maps does not affect the results of asymmetry for H$\alpha$ and H$\beta$ maps. However, when estimating the asymmetry for $E(B-V)$ maps (Section \ref{subsec:a2nd}), the intrinsic value of $\mathrm{H}\alpha/\mathrm{H}\beta$ are different for star-forming spaxels and AGNs or LI(N)ERs. Therefore, we need to restrict the spaxels to star-forming ones. We retrieve these data from MaNGA using the Python toolkit Marvin \citep{cherinka2019}.

\subsection{Continuum maps}

The maps of continuum emission are constructed using synthetic photometry. We convolved MaNGA spectra with SDSS filter bandpass to obtain synthetic images of galaxies at each band. Since the wavelength range of MaNGA (3600{--}10300\angstrom) does not fully include the SDSS $u$ and $z$ bands, we only present the results of $g$, $r$, and $i$ in this work. We tested our results using the SDSS photometry images and found that using synthetic photometry or the SDSS image data does not significantly affect our results. 

Instead of the image data from SDSS photometric observation, we use the synthetic method to avoid potential complications when matching the SDSS image data and the spectral data of MaNGA observations. The synthetic method ensures that the images match exactly with the MaNGA emission line maps pixel-by-pixel. The synthetic method is also more accurate and convenient to extract the rest-frame fluxes than applying the K-correction to the image data. Furthermore, MaNGA spectra have longer exposure time {(3 hours)} compared to SDSS images {(54 second)}. 
 MaNGA spectrum reaches a signal-to-noise ratio of 5 \AA$^{-1}$ fiber$^{-1}$ in the $r$-band continuum at a Galactic extinction corrected $r$-band surface brightness of 23 mag arcsec$^{-2}$ \citep{blanton2017}, while SDSS images only have a 5$\sigma$-depth of 22.7 mag at $r$-band. Therefore, the synthetic images have higher S/N than SDSS images.

\section{Method}
\label{sec:method}

\subsection{Asymmetry parameters}

\begin{figure}
    \centering
    \includegraphics[width=9cm]{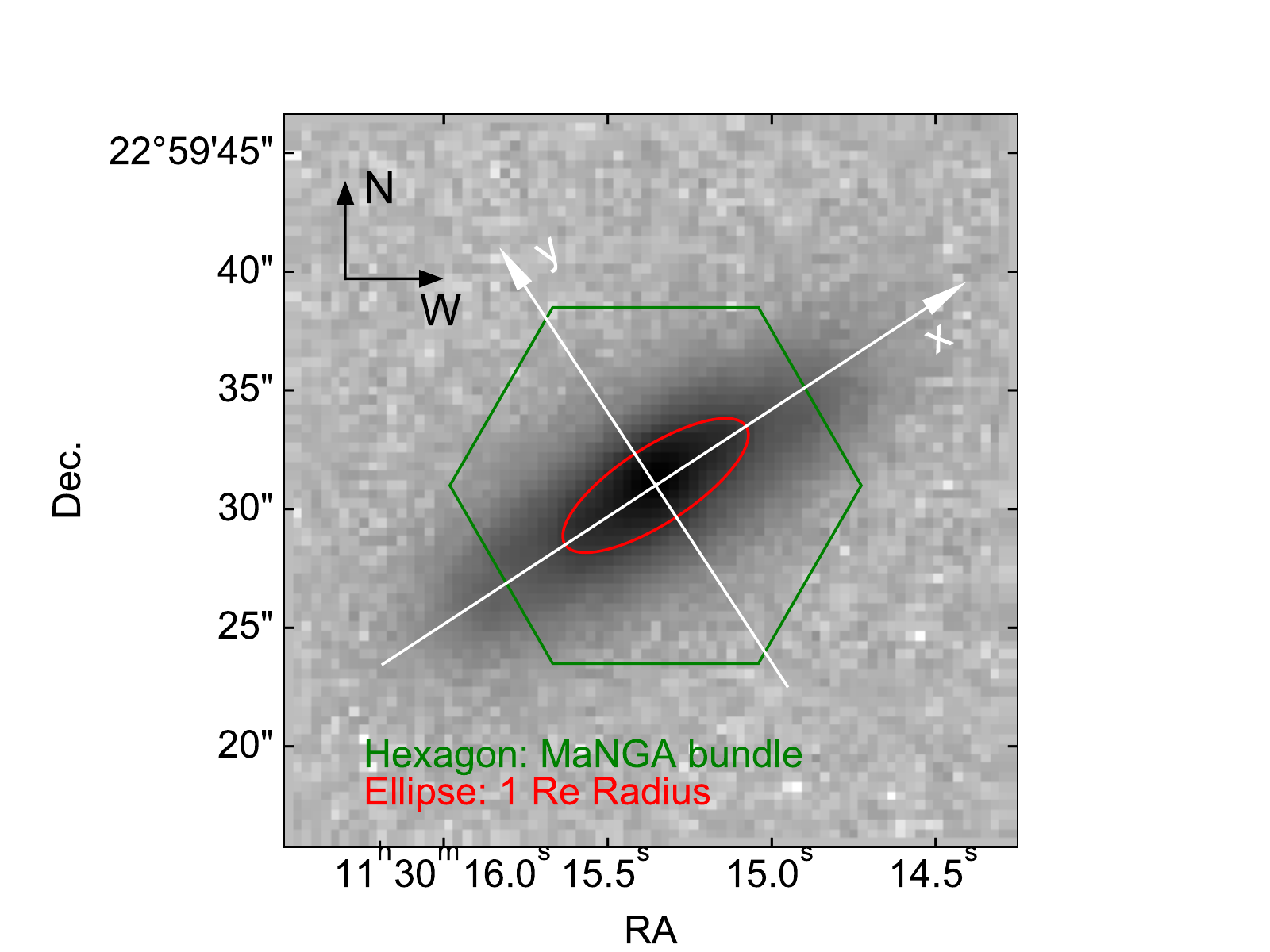}
    \caption{Illustration of dividing a galaxy image into halves according to the major and minor axes.{The green hexagon indicates the MaNGA bundle that covers this galaxy. The red ellipse shows the 1~$R_e$ radius of this galaxy.}}
    \label{fig:image}
\end{figure}


We introduce a new asymmetry parameter adopting the method used by \citet{walterbos1988} (see their Equation 4). 
By dividing the galaxy image into two halves by the major axis, we define the asymmetry about the major axis of a galaxy, $A_a$, as follows:
\begin{equation}
    \label{equ:ud}
   A_a \equiv {-2.5\log\left(\frac{I_\mathrm{upper}}{I_\mathrm{lower}}\right)} 
\end{equation}
where $I_\mathrm{upper}$ and $I_\mathrm{lower}$ are the mean surface brightness of the north and south half of the galaxy's image divided by the major axis, respectively. We calculate $I_\mathrm{upper}$ and $I_\mathrm{lower}$ within $1$ $R_e$ for each galaxy {as shown in Figure \ref{fig:image}. We choose $1$ $R_e$ to ensure the coverage of MaNGA observation and the S/N of the spaxels.} When dust is present in a galaxy, $A_a$ may vary at different inclination for observers.

Similarly, we calculate the asymmetry about the minor axis for comparison:
\begin{equation}
    \label{equ:lr}
   A_b \equiv {-2.5\log\left(\frac{I_\mathrm{left}}{I_\mathrm{right}}\right)}. 
\end{equation}
In contrast with $A_a$, the asymmetry about the minor axis $A_b$ should not be correlated with dust attenuation and thus should not vary with galaxy inclination.  {The $A_a$ and $A_b$ are applicable for both the continuum images and the emission line images. In the following, we use these parameters for our analysis.}


Unlike the classical asymmetry $A$, which uses the absolute values of the difference between two pixels, the asymmetry in Equations \ref{equ:ud} and \ref{equ:lr} calculates the difference between the two halves of the galaxy image directly. In such a definition, the asymmetry of the background can be ignored because the random noise is reduced when subtracting one half of the image from the other half. In the following, we use $A'$ to denote the asymmetry defined using our method in order to distinguish it from the classical asymmetry. More specifically, $A'$ refers to $A_a$ and $A_b$.

Another difference between the asymmetry $A'$ and the classical asymmetry is that $A'$ has a sign. In a galaxy with symmetric structures, $A_b$ should be zero, while the sign of $A_a$ may indicate which side is the near side of the galaxy. Under the assumption of \citet{walterbos1988}, the side with brighter continuum emission should be the far side because most of the stars are in front of the thin dust layer in the galaxy mid-plane, as illustrated in Figure \ref{fig:illu}. Figure \ref{fig:illu} also shows that if the thick dust component has a significant enough optical depth, the sign of $A_a$ for the emission lines is opposite of that for the continuum. The sign may help us determine {how the disk of the galaxy tilts with respect of the plane of the sky} (Section \ref{subsec:nearfar}).


Statistically, the sign of $A_b$ should be random for each galaxy because the structural asymmetry presents randomly in different galaxies. The sign of the asymmetry $A_a$ should also be random because galaxy disks tilt randomly in the space. Therefore, we expect the average value of the asymmetries $A'$ (both $A_b$ and $A_a$) to be zero. 
The average of the absolute value, $|A'|$, can be used to examine the dust asymmetry of galaxies. We note that $|A_a|$ equals to the $A_{\lambda}$ defined by \citet{walterbos1988}.

We divided the galaxy sample into five bins according to cosine of the inclination, $i$ (0.0-0.3, 0.3-0.45, 0.45-0.6, 0.6-0.75, 0.75-1). Using $b/a$ given by the NSA catalog gives similar results. Then we estimate the mean asymmetry in each bin and investigate the correlation between the inclination and asymmetries.

\subsection{Galaxy model}

Our primary aim of this work is to examine the asymmetry caused by dust in galaxies and explore its implication on dust-star geometry and galaxy opacity. To understand the results of the asymmetries, we use a simple but realistic galaxy model for comparison. The model is based on the galaxy models that have been proven to reproduce a realistic geometry of dust and stars in edge-on galaxies \citep[e.g.,][]{kb1987,xilouris1999,popescu2000, tuffs2004,bianchi2007,baes2010}.



The model is illustrated in Figure \ref{fig:illu}. It consists of a dustless bulge (not shown in the figure) and an exponential disk of stars, a thick dust component, and a thin dust component. We assume that the recombination lines originate from the clumpy components located in the thin disk. 
We do not distinguish the clumpy dust in the birth clouds and the diffuse dust in the thin disk but consider only the effective optical depth of the thin disk. 

The parameters of the stellar disk include a scale length $h_s$ and a scale height $z_s$. The thin dust disk is parametrized with a scale length $h_d$, a scale height $z_d$, and a face-on optical depth $\tau_V$. The thick dust disk is parametrized with a scale length $h_d^t$, a scale height $z_d^t$, and a face-on optical depth $\tau_V^t$. According to previous studies based on edge-on galaxies, $h_s\sim h_d \sim h_d^{t}$, $z_d < z_d^t\sim z_s$, $\tau_V>\tau_V^t$. To qualitatively compare with the observations, here we adopt the values of these parameters from \citet{tuffs2004} and \citet{chevallard2013}. The components and their parameters are listed in Table \ref{tab:model}.

We use the Monte Carlo dust radiative transfer code SKIRT \citep{baes2011,campsbaes2015} to simulate images based on this model. We use the monochromatic simulation for each band and do not include the information of the stellar population in the simulation to focus on the geometric parameters and optical depths. The images of different bands are different due to the various optical depths set for each band. Assuming the Calzetti attenuation curve \citep{calzetti2000}, we obtain that $\tau_g=1.2\tau_V$, $\tau_r=0.87\tau_V$, $\tau_i=0.68\tau_V$. The bulge to total luminosity ratio (B/T) can also change at different bands. Here we fix the B/T at each band to 0.2. 
The resolution of the images is set to 1 kpc to simulate the images obtained by MaNGA \footnote{We tested the resolution from 0.4 kpc to 1 kpc and found that it does not affect our results.}. 

 This geometric model only considers the effect of the dust attenuation and does not include any radiative transfer process of the gas. We simulate the line emissions separately, assuming a different geometry in which the line emissions are associated with the thin dust disk. A more detailed discussion can be found in Section \ref{sec:dustgeo}.

We then calculate the $A_a$ of these images in the same way we have done for the MaNGA data. {In the model, we do not introduce any structural asymmetry, such as spiral arm or clumpy region. The asymmetry about the major axis of the galaxy ($A_a$) is purely due to the observational effect caused by the thin dust disk when viewing the galaxy from an inclined angle.}  
The symmetric structure also results in that $A_b$ is zero for the simulated images. We compare our results with the results of the model in Section \ref{sec:dustgeo}. We focus on investigating the relations of the scale height and the optical depths of these components inferred from our results.


\setlength{\textfloatsep}{0.8cm}
\begin{table}
    \centering
    \caption{Galaxy model parameters for the stellar and dust components. For all the disk parameters, $h$ represents the scale length, $z$ the scale height, $\tau_V$ the face-on optical depth at V-band. For the bulge, $Re$ is the effective radius, $q$ the flattening factor, and B/T the bulge to total luminosity ratio.}
    \begin{tabular}{lrr}
    \hline
        Component & Parameter & value\\
    \hline
        Stellar disk & $h_s$ & 4200 pc\\
                    &  $z_s$ & 220 pc\\
        \hline
        Thin dust  disk & $h_d$ & 3000 pc \\
                    & $z_d$ & 48 pc\\
                    & $\tau_V$ & 0.1, 0.2, 0.4\\
        \hline
        Thick dust disk & $h_d^t$ & 4200 pc\\
                    &  $z_d^t$ & 140 pc\\
                    & $\tau_{V}^t$ & $1/3\tau_V$\\
        \hline
        Bulge  & $Re$ & 690 pc\\
            & $q$ & 0.7\\
            & B/T & 0.2\\
    \hline
    \end{tabular}
    
    \label{tab:model}
\end{table}

\section{Results}
\label{sec:results}

\subsection{Asymmetry of broadband images}
\label{subsec:acon}
\begin{figure*}
    \centering
    \includegraphics[width=18cm]{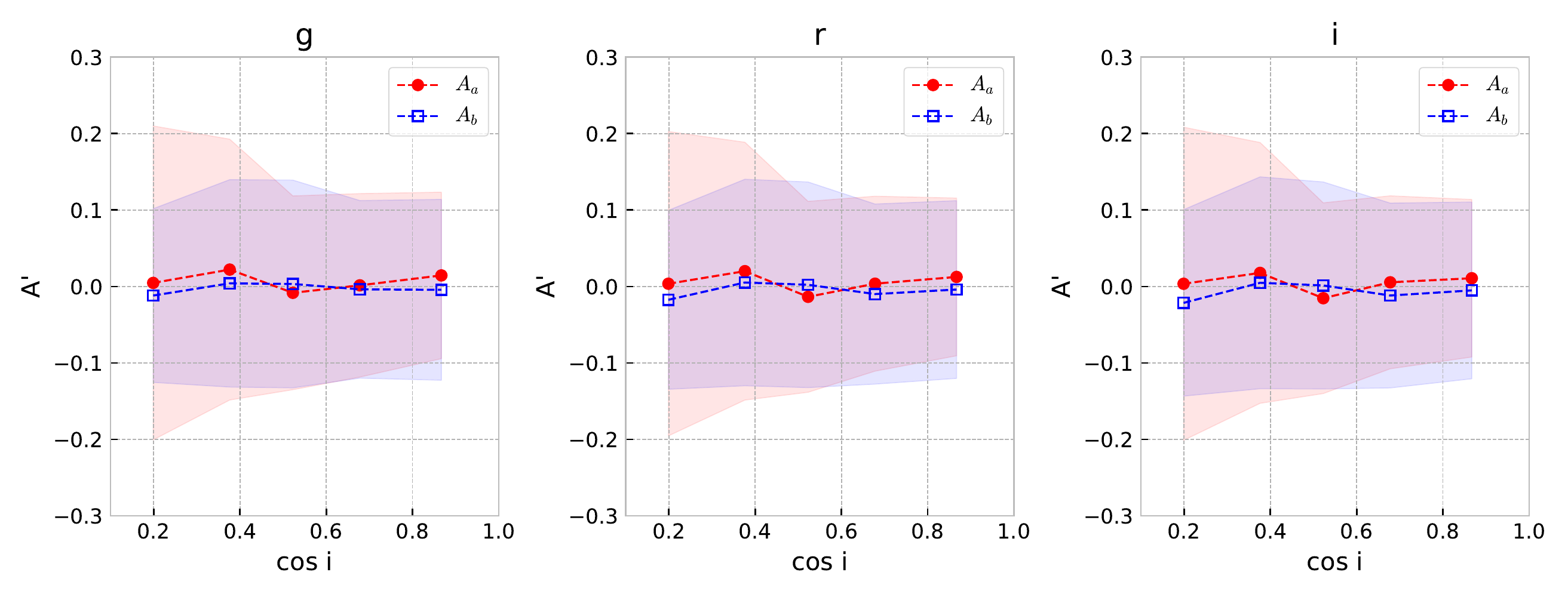}
    \caption{Asymmetries of galaxy images about its major ($A_a$, red) and minor axes ($A_b$, blue) as a function of inclination $i$. The shaded regions represent the standard deviations for galaxies in each bin.}
    \label{fig:synphot_res}
\end{figure*}

First, we test the asymmetry about the major and minor axes of broadband images. Figure \ref{fig:synphot_res} shows that the average of these asymmetries at each inclination bin is about zero, as predicted in Section \ref{sec:method}, because galaxies tilt randomly in space. Figure \ref{fig:synphot_res} also confirms that our method does not introduce any bias to any side of the galaxy image.

\begin{figure*}
    \centering
    \includegraphics[width=18cm]{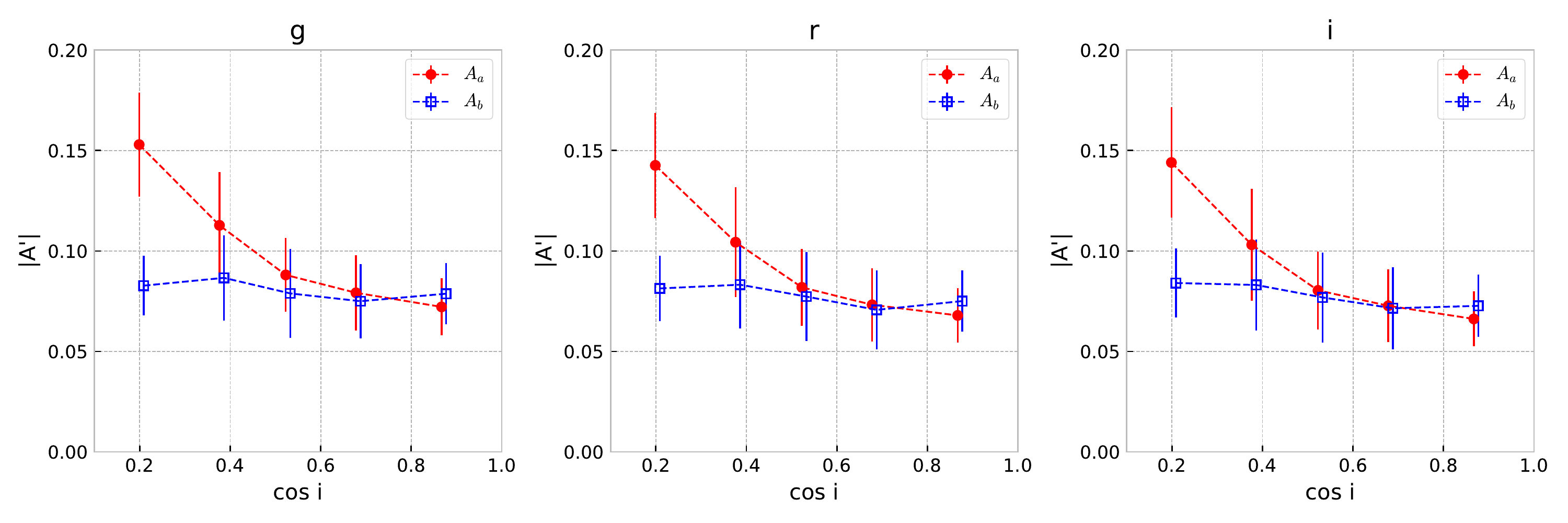}
    \caption{Asymmetries of galaxy images about its major ($|A_a|$, red) and minor axes ($|A_b|$, blue) as a function of inclination $i$. The error bars represent the $3\sigma$ error for galaxies in each bin, where $\sigma$ is the standard error of the mean. The standard error $\sigma=s/\sqrt{n}$, where $s$ is the standard deviation of the sample, and $n$ the sample size. }
    \label{fig:synphot_res_abs}
\end{figure*}

Figure \ref{fig:synphot_res_abs} shows the absolute values of these asymmetries.  The asymmetry about the minor axis $A_b$ is almost a constant as the inclination increases, i.e., there is no dependence of $|A_b|$ on inclination. However, for the asymmetry about the major axis, there is a trend that at a larger inclination, $|A_a|$ is larger, implying that the broadband fluxes of the highly inclined galaxies are affected more significantly by dust than the face-on galaxies. The results are consistent with the galaxy model with a thin disk of dust in the mid-plane. As illustrated in Figure \ref{fig:illu}, this component of dust can cause the asymmetry of continuum fluxes at the near and far side of galaxies. The difference between the edge-on bin ($\cos i < 0.3$) and the face-on bin ($\cos i > 0.75$) is about 0.06 mag. 

\subsection{Asymmetry of line emissions}
\label{subsec:aem}
\begin{figure}
    \centering
    \includegraphics[width=0.85\linewidth]{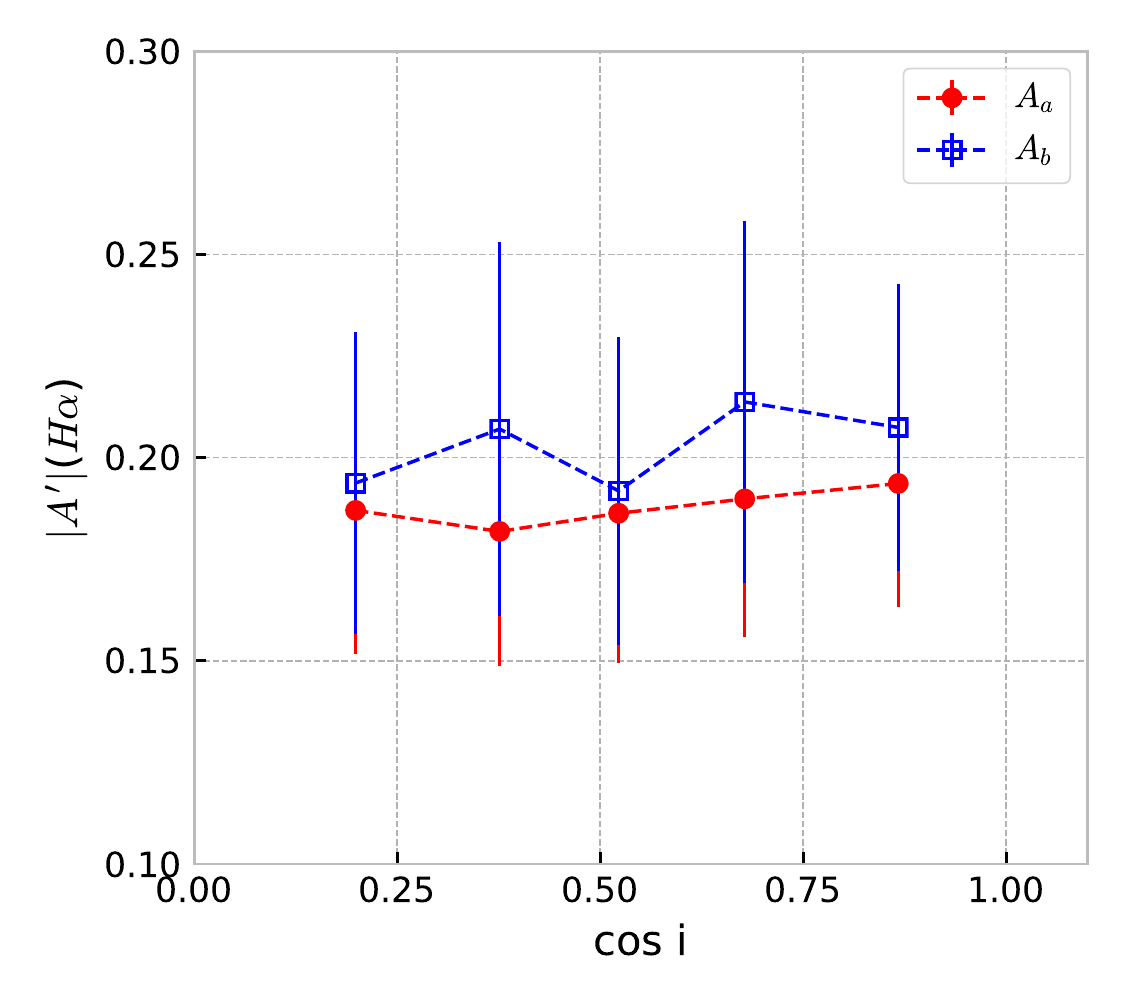}
    \caption{Asymmetries of H$\alpha$ maps about galaxies' major ($|A_a|$, red) and minor axes ($|A_b|$, blue) as a function of inclination $i$. The error bars represent the $3\sigma$ error for galaxies in each bin, where $\sigma$ is the standard error of the mean.}
    \label{fig:asym_ha}
\end{figure}

\begin{figure}
    \centering
    \includegraphics[width=0.85\linewidth]{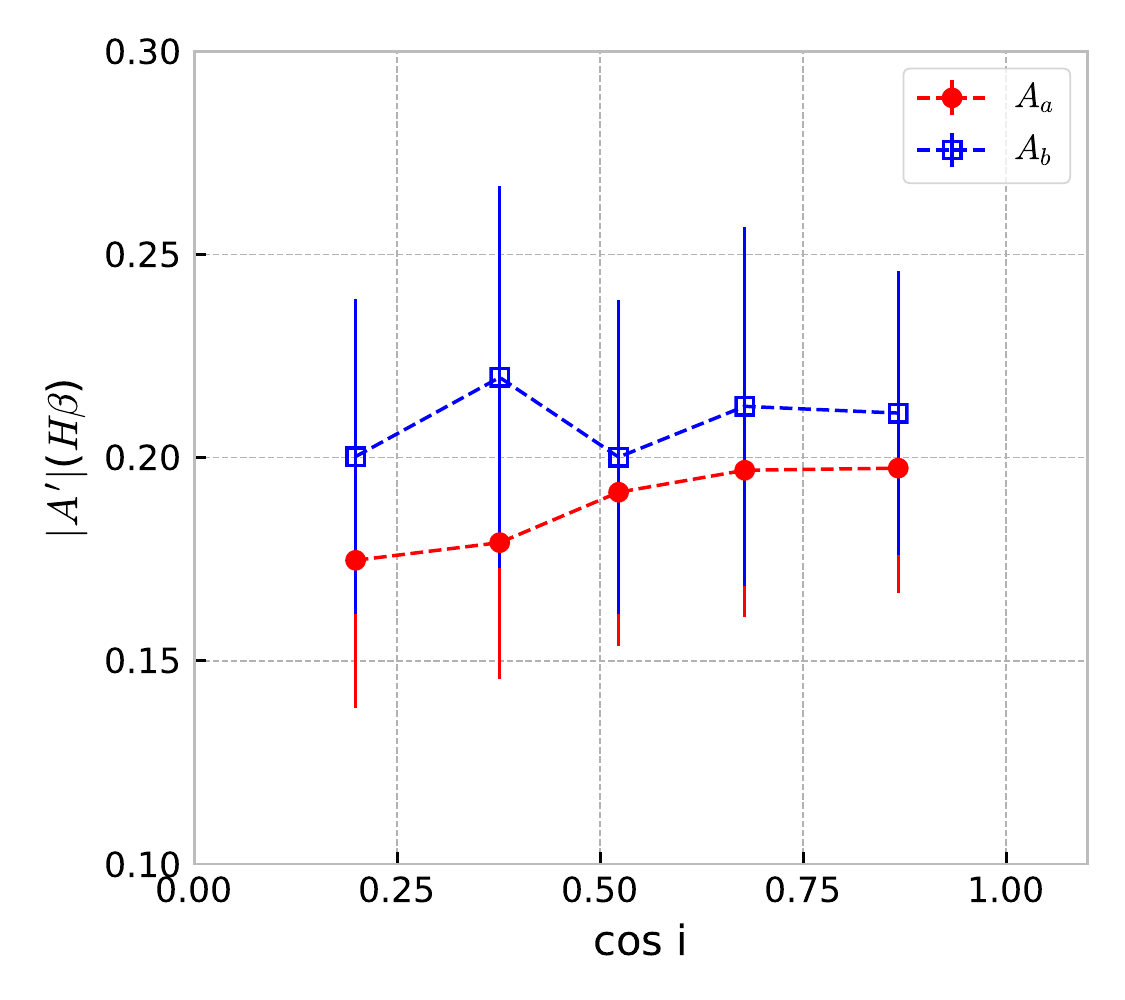}
    \caption{Asymmetries of H$\beta$ maps about galaxies' major ($|A_a|$, red) and minor axes ($|A_b|$, blue) as a function of inclination $i$. The error bars represent the $3\sigma$ error for galaxies in each bin, where $\sigma$ is the standard error  of the mean.}
    \label{fig:asym_hb}
\end{figure}

The asymmetries about the major and minor axes of the H$\alpha$ maps are shown in Figure \ref{fig:asym_ha}. On the contrary to the broadband images, there is no apparent trend for either $|A_a|$ or $|A_b|$ as the inclination increases. We also estimated the asymmetry of the H$\beta$ maps. H$\beta$ is at a shorter wavelength and, therefore, should be affected more by dust attenuation. However, the results are similar to those of the H$\alpha$ maps (Figure \ref{fig:asym_hb}). We also find that $|A_b|$ of the emission lines surpasses $|A_a|$ at some inclination bins. We will discuss the possible cause in Section \ref{subsec:impem}.

We note that for the emission lines, $|A_b|$ is around 0.2, which is {larger} than the value for the continuum images ($\sim 0.1$). The possible reason is that the $|A_b|$ values are correlated with the structural asymmetry of the galaxies, which is higher for clumpy emission-line regions than the stellar populations.


\subsection{Second order effects}
\label{subsec:a2nd}
\begin{figure*}
    \centering
    \includegraphics[width=18cm]{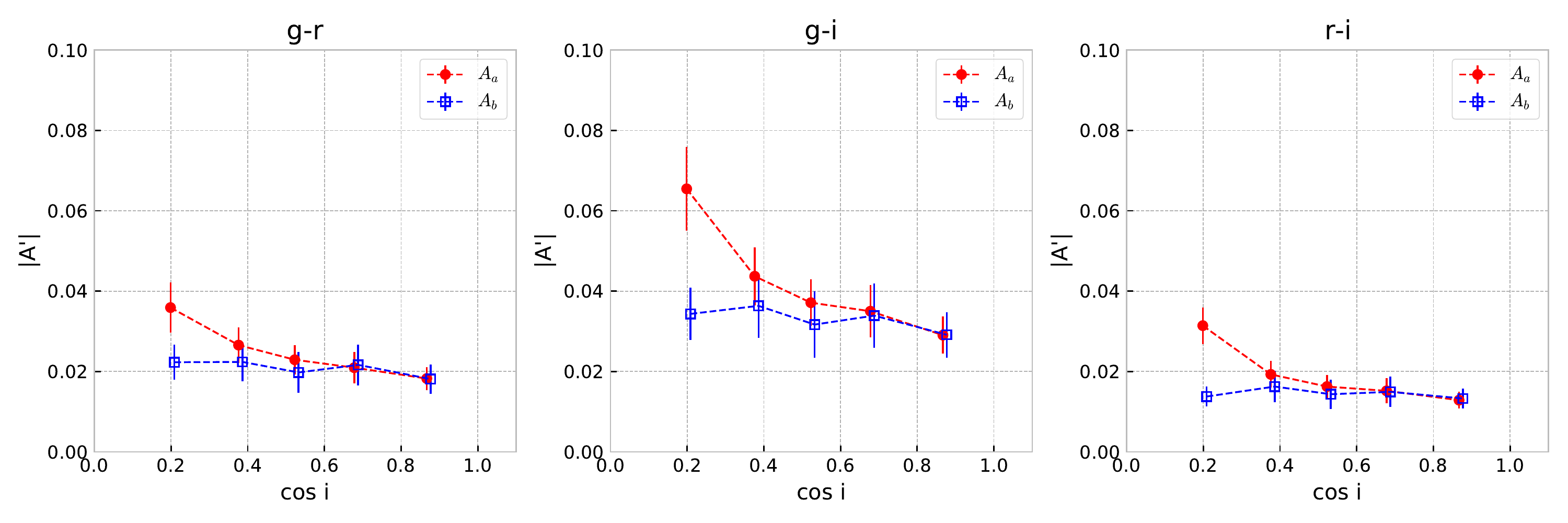}
    \caption{Asymmetries of \textit{g}-\textit{r} (\textit{left panel}), \textit{g}-\textit{i} (\textit{middle panel}), and \textit{r}-\textit{i} (\textit{right panel}) color maps about galaxies' major ($|A_a|$, red) and minor axes ($|A_b|$, blue) as a function of inclination $i$. The error bars represent the $3\sigma$ error for galaxies in each bin, where $\sigma$ is the standard error of the mean.}
    \label{fig:synphot_res_color}
\end{figure*}

In addition to the asymmetry of broadband and line emission maps, the dust obscuration can also cause the asymmetry for color maps because of the reddening effect. Therefore, we further examine the reddening effect by calculating the asymmetry of the color maps.  We find that for \textit{g}-\textit{i} colors, the asymmetry about the major axis $|A_a|$ of edge-on galaxies is apparently larger than that of the face-on galaxies, consistent with the geometry of a thin dust disk (Figure \ref{fig:synphot_res_color}). For the \textit{g}-\textit{r} and \textit{r}-\textit{i} colors, $|A_a|$ of edge-on galaxies is also larger compared to the face-on galaxies, but the difference is not as significant as \textit{g}-\textit{i} colors. It is conceivable because the difference between the wavelengths of \textit{g} and \textit{i} bands is larger than that of \textit{g} and \textit{r} or \textit{r} and \textit{i} bands. Therefore, the reddening effect is more apparent for \textit{g}-\textit{i} colors. The asymmetry $|A_b|$ of all these colors is constant for galaxies at different inclination. 

For emission lines, we investigate the asymmetry of the $E(B-V)_g$ map. The color excess of nebular emissions $E(B-V)_g$ is derived from
\begin{equation}
    E(B-V)_g=1.97\log \left[\frac{\mathrm{(H\alpha/H\beta)_{obs}}}{2.86}\right],
\end{equation}
assuming the reddening curve of \citet{calzetti2000}. The factor 2.86 is the intrinsic ratio of H$\alpha$ to H$\beta$, assuming that the ions undergo case B recombination \citep{of2006}. The value is applicable only for star-forming regions. 

\begin{figure}
    \centering
    \includegraphics[width=0.85\linewidth]{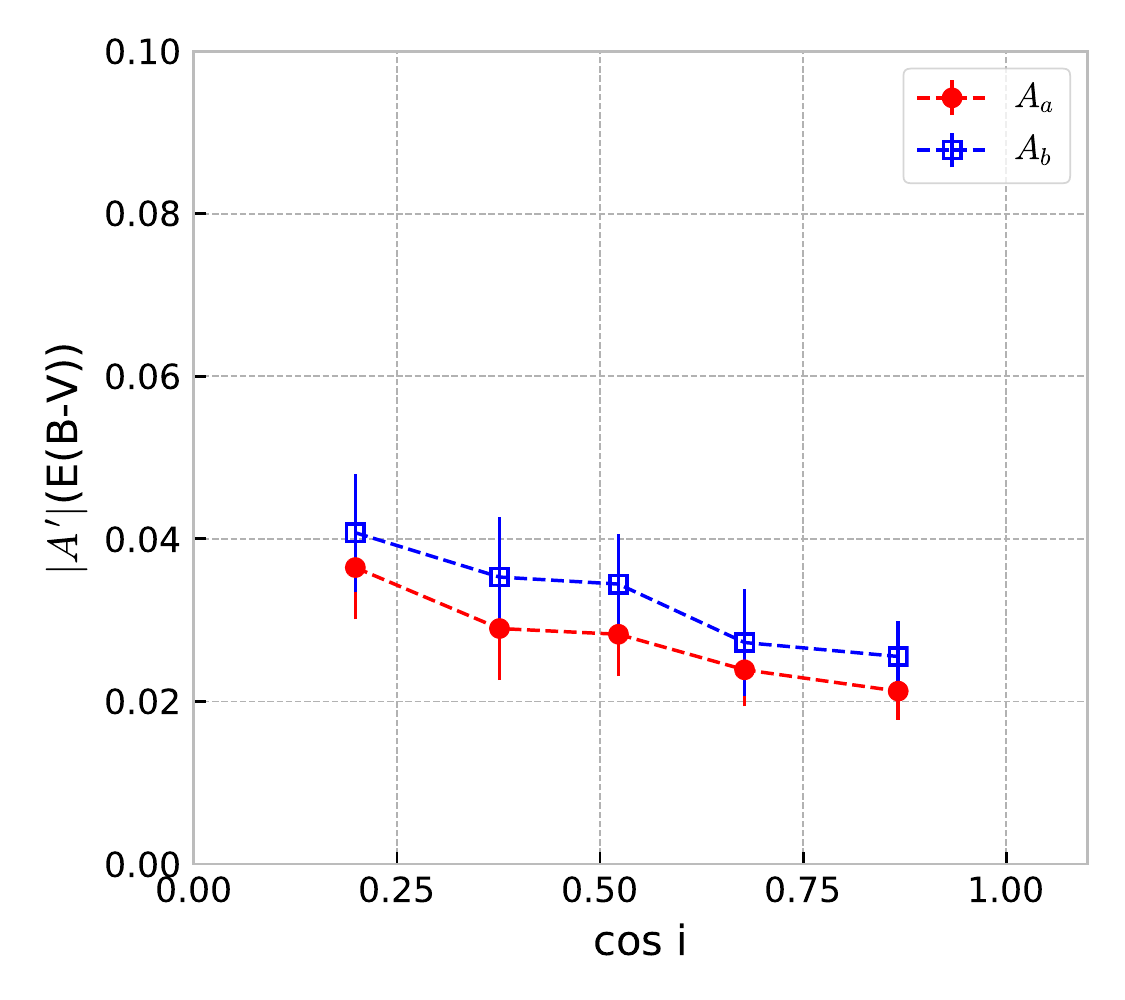}
    \caption{Asymmetries of $E(B-V)_g$ maps about galaxies' major ($|A_a|$, red) and minor axes ($|A_b|$, blue) as a function of inclination $i$. The error bars represent the $3\sigma$ error for galaxies in each bin, where $\sigma$ is the standard error of the mean.}
    \label{fig:ebvg}
\end{figure}

The results are shown in Figure \ref{fig:ebvg}. We find that $|A_a|$ and $|A_b|$ behave similarly. There is no apparent trend with inclinations for the difference between $|A_a|$ and $|A_b|$. The small decrease ($<$0.015) from the edge-on to face-on bins is due to the fact that the dispersion of the spaxels inside each galaxy increases about 0.03 on average from face-on to edge-on because there are less spaxels inside $1R_e$ for edge-on galaxies. The large dispersion causes the average of the asymmetry scatter to relatively larger values.

\section{Implications for dust-star geometry and galaxy opacity}
\label{sec:dustgeo}

In summary, by examining the asymmetry about the galaxy's major and minor axes ($|A_a|$ and $|A_b|$), we obtain the following results:

(1) For the continuum fluxes, the asymmetry about the major axis ($|A_a|$) increases with the inclination. When galaxies become close to edge-on, this asymmetry becomes more evident. On the contrary, the asymmetry about the minor axis ($|A_b|$) is a constant as inclination changes.

(2) For the emission line fluxes, that is to say, H$\alpha$ and H$\beta$, neither $|A_a|$ nor $|A_b|$ show significant correlation with inclination. 

(3) We also examine the asymmetry of the colors derived from continuum fluxes and find that the $|A_a|$ values of $g-r$, $g-i$ and $r-i$ colors are apparently larger for highly inclined galaxies, whereas no trend with inclination is found concerning the $|A_b|$ ones.

(4) By calculating $E(B-V)_g$ from the H$\alpha$ to H$\beta$ line ratio, we find that $|A_a|$ and $|A_b|$ are quite similar. The difference does not depend on inclination.

For the continuum image, we find that $|A_a|>|A_b|$ for galaxies with inclination $\cos i < 0.5$, suggesting that the asymmetry caused by dust is statistically detectable for galaxies at such inclinations. For galaxies with lower inclination ($\cos i > 0.5$), we do not find detectable asymmetry caused by dust. The results imply that the asymmetry caused by dust is a potential method to determine the near and far side of galaxies. We further discuss the effect in Section \ref{sec:discussion}.
In this section, we discuss the constraints that our results put on the dust-star geometry and {galaxy opacity.}
\subsection{Implications of broadband image asymmetry}
\label{subsec:impcon}

Our results of the broadband images can be explained by models in which the scale height of the dust component is smaller than that of the stellar component. As described in Section \ref{sec:method}, the galaxy model we adopt has a thin dust disk with a scale height $z_d=48$ pc, much smaller than the scale height of the stellar disk ($z_s=220$ pc) and the bulge $R_e$ ($690$ pc).

\begin{figure*}
    \centering
    \includegraphics[width=18cm]{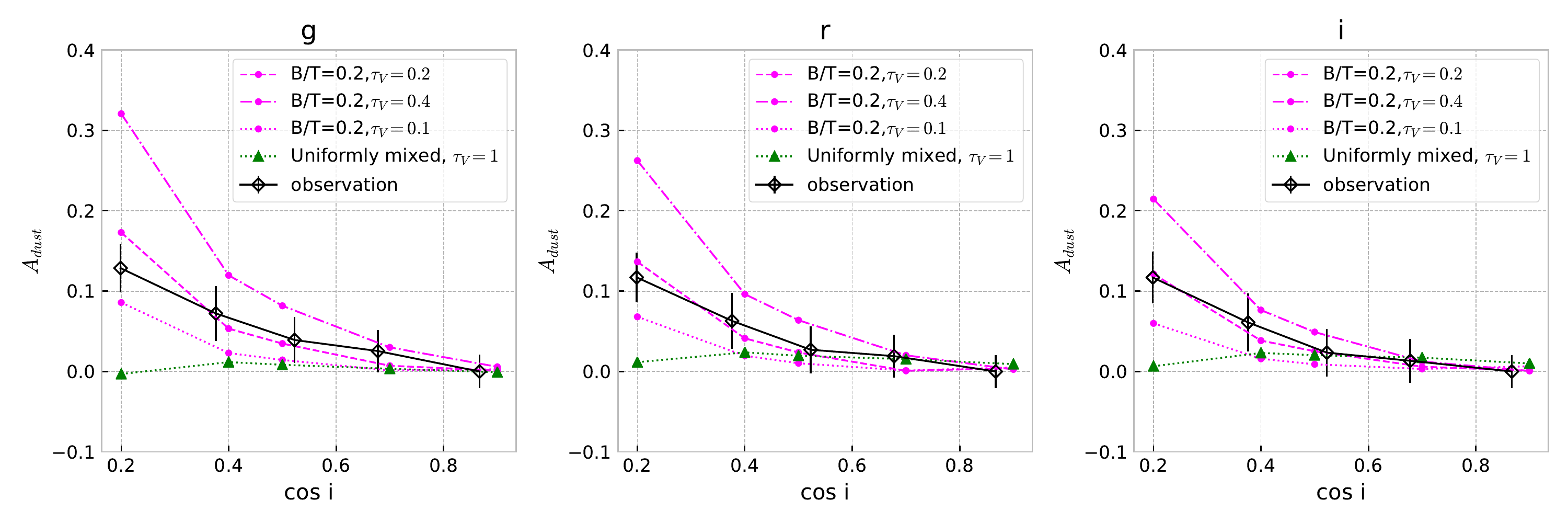}
    \caption{Dependence of $A_\mathrm{dust}$ on inclination. The diamonds with error bars and solid lines indicate the observational results. Magenta dots are the SKIRT simulation results for the galaxy model described in Section \ref{sec:method} with B/T$=0.2$ and the thin dust disk optical depth $\tau_V=0.1$ (dotted lines), $0.2$ (dashed lines), and $0.4$ (dash-dotted lines). Green triangles are the SKIRT simulation results for a disk of uniformly mixed stars and dust with $\tau_V=1$.}
    \label{fig:skirt}
\end{figure*}

Based on this model, we can quantitatively compare our results with those from SKIRT simulation to constrain the optical depth of the dust components. The images from the simulation do not have any structural asymmetry. Therefore, $|A_a|$ of the simulated image is the asymmetry caused by dust, $A_\mathrm{dust}$. In observations, on the other hand, $|A_a|$ includes both the structural asymmetry, $A_\mathrm{struc}$ and the dust asymmetry $A_\mathrm{dust}$. Here we assume that $|A_b|=A_\mathrm{struc}$ and $A_a^2=A_\mathrm{dust}^2+A_\mathrm{struc}^2$, and we can obtain the value of $A_\mathrm{dust}$ for our sample at each inclination, as shown in Figure \ref{fig:skirt}. 

The results based on the SKIRT simulation are overplotted in Figure \ref{fig:skirt}. As listed in Table \ref{tab:model}, we assume that a B/T$=0.2$ and $\tau_V=0.1$, $0.2$ and $0.4$. We fixed the optical depth of the thick dust component ($\tau_V^t$) to $1/3\tau_V$. Comparing the asymmetry obtained from the SKIRT simulation and the MaNGA observation, we find that the dependence of the $A_\mathrm{dust}$ on the inclination of the sample is consistent with a model with a $\tau_V\sim0.2$. If the dust and stars are uniformly mixed, $A_\mathrm{dust}$ is about zero for all inclinations, as shown in Figure \ref{fig:skirt}.



The asymmetry we find for the continuum light can be caused by both the thin, diffuse dust in the disk and the clumpy birth-cloud dust. However, if the latter dominates, the covering factor $F$ of the birth clouds needs to be high, which is in contrast with what has been found in previous works \citep[e.g.,][]{wild2011}. The results of the colors also support that $F$ should be small (Figure \ref{fig:synphot_res_color}). If the asymmetry is caused mainly by the clumpy components which are optically thick ($\tau_\mathrm{BC}>>1$), all light behind the dust should be absorbed. Even if we could still see the correlation of the broadband flux asymmetry with inclination, there would be no reddening effect. Therefore, the trend of the colors with inclination shown in Figure \ref{fig:synphot_res_color} favors a thin diffuse dust component with an intermediate $\tau$.

The existence of the dustless bulge brings more complications in our study. Without the thin dust component, the thick dust component can cause similar asymmetries for galaxies because the bulge has a $R_e$ (690 pc) larger than the scale height of the thick dust component ($z_d^t=140$ pc). However, if the thick dust component causes the asymmetries, $\tau_V^t$ must be $\sim 0.4$, twice larger than that is required for the thin dust component. 

The bulge to total luminosity ratio (B/T) affects the results significantly. As B/T increases, the same $\tau_V$ can cause larger asymmetry. According to \citet{simard2011}'s results, the average B/T of our sample is about 0.13 at $g$-band, and 0.16 at $r$-band. However, according to \citet{meert2015,meert2016}'s results, the average B/T is 0.18, 0.18 and 0.2 for $g$, $r$, and $i$ bands, respectively. In Figure \ref{fig:skirt}, it appears that the observed $g$-band asymmetry is slightly smaller than that in the $\tau_V=0.2$ model, consistent with a smaller B/T.

The uncertainties in B/T affect the color asymmetry. The asymmetry of the color map is a combined effect of the optical depth of the dust and the B/T at different bands. Moreover, the color asymmetry is also related to the attenuation curve, which is beyond the scope of this study.

\subsection{Implications of emission line asymmetry}
\label{subsec:impem}
For H$\alpha$ and H$\beta$, our results confirm that the scale height of the emission line regions is similar to the thin dust disk. If the emission lines are from a disk with a significantly larger scale height than the thin dust disk, we should have detected asymmetry for emission lines similar to that we find in broadband images. 

Assuming that the emission line regions are uniformly distributed in the thin disk, the thick dust component can also cause the asymmetry for the line emission because the path lengths through the thick dust component become different for the near and far side of galaxies when the galaxies tilt (Figure \ref{fig:illu}).  However, our results of emission line asymmetry and $E(B-V)_g$ asymmetry do not show any evidence for the existence of the thick component. From the SKIRT simulation, we also find that if the emitting source is in the thin disk (i.e., $z_s=z_d$, $h_s=h_d$), the obtained asymmetry for the galaxies in the largest inclination bin ($\cos i<0.3$) is less than 0.04 even for a relatively large $\tau$ ($\tau_V=1.0$, $\tau_V^t=0.33$). 

The results can be explained by the fact that the emission lines are predominantly attenuated by the birth clouds from which they originate. The different opacity caused by the uneven path-length through the diffuse ISM dust in the thick component is insignificant compared to the dust opacity difference between the birth clouds and ISM. The results are consistent with that found by previous studies such as \citet{yip2010} and \citet{wild2011}. The recent works of \citet{mingozzi2020} and \citet{greener2020} show that the dust extinction in the gas has radial profiles that are low and flat for low mass galaxies and tend to increase and steepen with the stellar mass, suggesting a geometry in which both thin ISM dust disk and clumpy dust are distributed with a smaller scale height than the old stellar populations. Their findings also agree with our results that the gas emission is mainly attenuated by the thin dust components. 

In our study, the results can also be affected by other factors. For instance, the line emission, different from the continuum emission, is distributed irregularly in galaxies. Therefore, the intrinsic asymmetry of the emission line maps is quite high. Even if there is some asymmetry caused by ISM dust, it is insignificant compared to the intrinsic asymmetry. 


For our emission line results, it is intriguing that $|A_b|$ sometimes is higher than $|A_a|$ although the difference is within the $3\sigma$ error. The higher value of $|A_b|$ can be caused by the random distribution of HII regions in galaxies. Another possibility is that the difference is related to the structure of spiral arms. For example, when a galaxy inclines to a specific position, the inner and outer parts of arms can be seen respectively on the left and right of the images, causing the increase of the asymmetry about the minor axis at that inclination.   


In summary, the results rule out the simple model of a uniform and homogeneous layer of dust and stars. The asymmetries we found support the existence of a thin dust layer in the galaxy mid-plane with $\tau_V\sim0.2$, consistent with the results of \citet{shao2007}, who found that the face-on galaxies are on average 0.2 mag brighter than inclined galaxies. Our results also suggest that the thick dust component is not significant to cause detectable asymmetry for H$\alpha$ and H$\beta$ lines, consistent with what has been found in \citet{yip2010, wild2011}, and a recent work of Lu et al. (in prep.).

\section{Discussion}
\label{sec:discussion}

\subsection{Dust asymmetry and structural asymmetry}
\label{subsec:discuss_asymmetry}

The classical asymmetry parameter, $A$, is commonly used to quantify galaxy morphology and to identify galaxy mergers \citep[e.g.,][]{conselice2000,lotz2008}. It is defined as the difference between the image and its 180-degree rotation:
\begin{equation}
A=\frac{\sum_{i,j}|I_0(i,j)-I_{180}(i,j)|}{\sum_{i,j}|I_0(i,j)|}-B_{180},
\end{equation}
where $I_0$ is the galaxy's image and $I_{180}$ is the image rotated by 180 degrees about the galaxy's central pixel, and $B_{180}$ is the mean asymmetry of the background. 

The classical asymmetry parameter calculates the difference between the image $I_0$ and its 180-degree rotation $I_{180}$. More specifically, the asymmetry $A$ measures the width of the flux distribution of the residual image $I_0-I_{180}$. The nature of $I_0-I_{180}$ residual image determines that we can not extract more information other than the distribution width, because the mean of any $I_0-I_{180}$ residual image is zero. $A$ cannot discriminate the case that one half of the galaxy is brighter than the other half and the case that the asymmetric patterns are randomly distributed in two sides of the galaxy. However, by examining half of the $I_0-I_{180}$ residual image, we can obtain more information. Our method is equivalent to taking half of the $I_0-I_{180}$ residual image and examine its mean value. 

In short, the classical asymmetry contains not only the structural asymmetry but also the dust asymmetry. Even if the galaxy structure is strictly symmetric, we can observe asymmetry when the galaxy inclines due to the dust attenuation. Our results show that the asymmetry due to the dust attenuation is detectable when the galaxies are close to edge-on.
Therefore, the asymmetry measured by $A$ overestimates the structural asymmetry for highly inclined galaxies. To obtain precise measurements of structural asymmetry, we should first correct the dust attenuation for each galaxy. Similar results have also been found by a recent work of \citet{li2020}. Taking advantage of the MaNGA data and their dust correction method,  they find that the asymmetric profile for highly inclined galaxies becomes symmetric after correcting for the dust attenuation.

\subsection{Implication of galaxy tilt}
\label{subsec:nearfar}


The asymmetry caused by dust implies that it can be used to determine the near side of a galaxy, which should have a fainter and redder optical continuum than the far side. Several works have used the asymmetric profiles to determine the near and far side of galaxies \citep[e.g.,][]{devaucouleurs1958,buta2003}. However, these works are limited to a few nearby galaxies and include detailed analyses of galaxies' light profiles, which are hardly applicable for larger samples or more distant galaxies. 

If there is also a detectable asymmetry of emission lines caused by the thick ISM dust component, the nearer side should also have a brighter line emission. The combination of the asymmetries of the continuum and line emission thus provides strong constraints on galaxy tilt. We found that in some galaxies, the continuum image and the emission line image do have opposite asymmetries (e.g., MaNGA-ID 1-491233). Unfortunately, our results show that the asymmetry of line emissions is not detectable for most galaxies, and the dust asymmetry is not significant enough to determine the tilt for individual galaxies.






Can dust asymmetry be used as an indicator of galaxy tilt? Even without considering the more complicated radiative transfer processes in stars and dust (such as the forward scattering, or more complicated geometry of stars and dust), it is still challenging. First, whether the dust asymmetry is a reliable indicator of galaxy tilt is hard to test. An optional method (maybe the only other method) to determine the near- and far-side of galaxies is using spiral arms. Assuming all spiral arms are trailing, and combining the kinematic maps, one can distinguish the near- and far-side of galaxies. This method has been used in several studies to determine the galaxy tilt and, therefore, the direction of gas flow \citep[e.g.,][]{ho2020}. The spiral arm method requires that the galaxy has prominent spiral arms, which means that the stars and dust in that galaxy are not distributed in pure exponential form as we assumed. Also, the spiral arms are more visible when face-on, while we need a high inclination to detect the dust asymmetry. Therefore, the two methods are for different types of galaxies and hardly comparable. Furthermore, the assumption that the spiral arms are all trailing is not valid. In the nearby universe, several galaxies have been proved to have leading spiral arms \citep{buta2003}.     

Second, the effectiveness of the dust asymmetry method is limited. When galaxies become smoother (the structural asymmetry to be small), the dust asymmetry method should be more effective. However, in galaxy evolution, the smoother galaxies contain more evolved stellar populations and less dust to cause significant dust asymmetry. 

However, we cannot rule out the potential of using dust asymmetry to determine the galaxy tilt. Our results show that the difference between $A_a$ and $A_b$ is about 0.06 mag when $\cos i\sim 0.2$. However, it is a statistical result and does not ensure that the asymmetry can be detected for an individual galaxy even if the measurement error is less than 0.06. Due to the dispersion of $A_a$ and $A_b$, it is quite possible that the difference between $A_a$ and $A_b$ is smaller than 0.06 mag for a galaxy. Therefore, a more detailed analysis is required to examine the asymmetry in a specific galaxy. With higher resolution images and more sophisticated methods in the future, the dust asymmetry may provide valuable information on the 3D information for distant galaxies.

\subsection{Uncertainties}
\label{subsec:uncertainties}

Similar to the computation of the classic rotational asymmetry, one of the most crucial aspects to calculate the dust asymmetry is the choice of a center. The centering problem has been discussed in almost all previous works on calculating morphological parameters. The ideal center should be the center of mass. Such a center, however, is difficult to measure accurately in observations. Traditionally, the center is determined either by using the brightest point \citep[e.g.,][]{abraham1994,abraham1996}, or an iterative process to find a minimum of the asymmetry parameter or the total second-order moment $M_\mathrm{tot}$ \citep[e.g.,][]{conselice2000,lotz2004,lotz2008,conselice2014}.

In this work, we chose the brightest point of $r$-band as the center for all our calculations considering the following reasons: First, as shown by \citet{li2020} in their Figure 9, the highly inclined galaxies have asymmetric light profiles due to the dust attenuation, and the profile would be symmetric after correcting for the dust effect. Therefore, if we use the asymmetry minimization, the center we found can be biased to the brighter side. Another reason is that we need to compare the dust asymmetry in different bands (including continuum bands, H$\alpha$, and H$\beta$ images), whereas the centers of asymmetry minimization at each band are not the same physical center. 

We tested the effect of using different centers by adding the line with the brightest pixels to the galaxies' fainter halves. In this way, we minimize the asymmetry obtained. We still find the $A_\mathrm{dust}$ for the continuum but smaller, consistent with a model with $\tau_V$ between 0.1 and 0.2. Note that the choice of the center causes the primary uncertainties for this work.

Our estimation based on Equation \ref{equ:ud} can be considered to be a lower limit of the real dust attenuation. The reasons have been given in \citet{walterbos1988}, including three facts that 1) the dust is not entirely in front of all the light, 2) light scattering has been ignored, and 3) although not significant, the far side is also obscured by dust.
Therefore, the real asymmetry caused by dust can be more significant than our results. Since  $|A_a|$ represents a lower limit, there are no tight correlations between the parameter $|A_a|$ and the dust attenuation estimated by other methods such as the SED fitting or the Balmer decrement.

For the asymmetry of line emission, other than the effects of the dust and intrinsic structural asymmetry, the $A_a$ and $A_b$ values can also be affected by the environment of the galaxy. As shown in the series of GASP and VESTIGE works \citep[e.g.,][]{poggianti2016,poggianti2017,moretti2020,boselli2018,fossati2018}, there are galaxies suffering from ram pressure stripping in galaxy clusters, resulting in that the H$\alpha$ images of these galaxies are quite asymmetric about its major or minor axes. \citet{poggianti2016} shows that these galaxies take $\sim 2$\% of star-forming galaxies. More galaxies could be affected by ram pressure stripping, considering that their sample only includes the most secure stripping candidates. Also, phenomena such as mergers or tidal interactions could play a role in producing asymmetry. According to previous studies\citep{ellison2008,li2008,pan2019,feng2019}, galaxies in pairs or multiples take about 5\% of the SDSS galaxy sample and therefore should not affect our statistical results. We tested the effect by removing the most asymmetric galaxies in our sample and find that our conclusion remains.

\section{Summary}
\label{sec:summary}
We selected a sample of 1320 galaxies from SDSS-MaNGA observation with stellar masses uniformly distributed in each $b/a$ bin ranging from $10^{9.5}~M_{\odot}$ to $10^{11}M_{\odot}$. We define the parameters $A_a$ and $A_b$ to estimate the asymmetries of a galaxy about its major and minor axes, respectively. We expect $|A_a|$ to reflect the asymmetry caused by dust attenuation. We examine these parameters for continuum and line images and their relations with galaxy inclination and find the following results:

1) We examine the asymmetry of the continuum fluxes and find that $|A_a|$ increases with the inclination, while $|A_b|$ is a constant as inclination changes. Similar trends are found for $g-r$, $g-i$ and $r-i$ color images.

2) For the H$\alpha$ and H$\beta$ images, neither $|A_a|$ nor $|A_b|$ shows a significant correlation with inclination. Also, by calculating $E(B-V)_g$ from H$\alpha$ and H$\beta$ line ratio, we do not find a significant dependence of the asymmetry of $E(B-V)_g$ on inclination.

Our results show that the asymmetry due to the dust attenuation is detectable when the inclination is larger than about $60^\circ$ ($\cos i < 0.5$). The results of the asymmetry about the major and minor axes imply that there is a dust component with a smaller scale height than that of the stars, consistent with the results of previous works based on observations and galaxy models \citep{xilouris1999,tuffs2004,popescu2011}. The emission line results show that statistically, the opacity caused by the diffuse ISM in the thick dust component is insignificant, consistent with the results of \citet{yip2010,wild2011}. Our results rule out the simple model that the dust and stars are uniformly mixed. Compared with the SKIRT simulation, our results suggest a galaxy model with the thin dust disk optical depth $\tau_V$ about $0.2$.

Our results suggest that the classical asymmetry parameter can overestimate the structural asymmetry of highly inclined galaxies because the parameter includes both the structural asymmetry and the asymmetry caused by the dust disk.   

The inclination and the dust attenuation are both complicated effects in observations and can cause many observational biases when measuring galaxy properties. Understanding better these effects further requires a larger sample of data with better resolution and higher qualities. In the study of \citet{buta2003}, the physical resolution of the image is $19.5$ pc, enabling a detailed analysis of the profiles of two sides of galaxies. The improvement of the measurement accuracy will also help in such studies. With such data, it is possible to use the inclination and dust effects to discriminate between the near and far side of a galaxy, which will provide an independent measurement for galaxy tilt other than the spiral arm method.

\acknowledgements
{We thank the anonymous referee for helpful and constructive comments that improve the paper. 
This work is partly supported by the National Natural Science Foundation of China (NSFC) under grant nos.11433003, 11573050, and by a China-Chile joint grant from CASSACA (PI: FTY and MB). MB acknowledges support from FONDECYT regular grant 1170618.

Funding for the Sloan Digital Sky Survey IV has been provided by the Alfred P. Sloan Foundation, the U.S. Department of Energy Office of Science, and the Participating Institutions. SDSS-IV acknowledges
support and resources from the Center for High-Performance Computing at
the University of Utah. The SDSS web site is www.sdss.org.

SDSS-IV is managed by the Astrophysical Research Consortium for the 
Participating Institutions of the SDSS Collaboration including the 
Brazilian Participation Group, the Carnegie Institution for Science, 
Carnegie Mellon University, the Chilean Participation Group, the French Participation Group, Harvard-Smithsonian Center for Astrophysics, 
Instituto de Astrof\'isica de Canarias, The Johns Hopkins University, Kavli Institute for the Physics and Mathematics of the Universe (IPMU) / 
University of Tokyo, the Korean Participation Group, Lawrence Berkeley National Laboratory, 
Leibniz Institut f\"ur Astrophysik Potsdam (AIP),  
Max-Planck-Institut f\"ur Astronomie (MPIA Heidelberg), 
Max-Planck-Institut f\"ur Astrophysik (MPA Garching), 
Max-Planck-Institut f\"ur Extraterrestrische Physik (MPE), 
National Astronomical Observatories of China, New Mexico State University, 
New York University, University of Notre Dame, 
Observat\'ario Nacional / MCTI, The Ohio State University, 
Pennsylvania State University, Shanghai Astronomical Observatory, 
United Kingdom Participation Group,
Universidad Nacional Aut\'onoma de M\'exico, University of Arizona, 
University of Colorado Boulder, University of Oxford, University of Portsmouth, 
University of Utah, University of Virginia, University of Washington, University of Wisconsin, 
Vanderbilt University, and Yale University.
}

\bibliography{ref}
\bibliographystyle{aasjournal}
\end{document}